\documentclass[11pt]{article}
\usepackage{amssymb,amsmath,amsfonts}
\usepackage{graphicx}
\usepackage{graphics}
\usepackage{eepic,epsfig}

\begin{document}

\title{Vacuum polarization by a global monopole with finite core}
\author{E. R. Bezerra de Mello$^{1}$\thanks{%
E-mail: emello@fisica.ufpb.br}\, and A. A. Saharian$^{1,2}$\thanks{%
E-mail: saharyan@server.physdep.r.am} \\
%EndAName
\\
\textit{$^1$Departamento de F\'{\i}sica-CCEN, Universidade Federal da Para%
\'{\i}ba}\\
\textit{58.059-970, Caixa Postal 5.008, Jo\~{a}o Pessoa, PB, Brazil}\vspace{%
0.3cm}\\
\textit{$^2$Department of Physics, Yerevan State University,}\\
\textit{375025 Yerevan, Armenia}}
\maketitle

\begin{abstract}
We investigate the effects of a $(D+1)$-dimensional global monopole core on
the behavior of a quantum massive scalar field with general curvature
coupling parameter. In the general case of the spherically symmetric static
core, formulae are derived for the Wightman function, for the vacuum
expectation values of the field square and the energy-momentum tensor in the
exterior region. These expectation values are presented as the sum of
point-like global monopole part and the core induced one. The asymptotic
behavior of the core induced vacuum densities is investigated at large
distances from the core, near the core and for small values of the solid
angle corresponding to strong gravitational fields. In particular, in the
latter case we show that the behavior of the vacuum densities is drastically
different for minimally and non-minimally coupled fields. As an application
of general results the flower-pot model for the monopole's core is
considered and the expectation values inside the core are evaluated.
\end{abstract}

\bigskip

{PACS number(s): 03.70.+k, 98.80.Cq, 11.10.Kk}

\newpage

\section{Introduction}

It is well known that different types of topological objects may have been
formed in the early universe after Planck time by the vacuum phase
transition \cite{Kibble,V-S}. Depending on the topology of the vacuum
manifold these are domain walls, strings, monopoles and textures. Among
them, cosmic strings and monopoles seem to be the best candidate to be
observed. A global monopole is a spherical heavy object formed in the phase
transition of a system composed by a self-coupling Goldstone field, whose
original global symmetry is spontaneously broken. The matter fields play the
role of an order parameter which outside the monopole's core acquires a
non-vanishing value. The global monopole was first introduced by Sokolov and
Starobinsky \cite{Soko77}. A few years later, the gravitational effects of
of the global monopole were considered in Ref. \cite{B-V}, where a solution
is presented which describes a global monopole at large radial distances.
The gravitational effects produced by this object may be approximated by a
solid angle deficit in the (3+1)-dimensional spacetime.

The nontrivial properties of the vacuum are among the most important
predictions of quantum field theory. These properties are manifested in the
response of the vacuum to the external electromagnetic and gravitational
fields. In particular, the explicit calculations of the vacuum polarization
caused by particular external fields have played an important role in the
development of quantum field theory. The quantum effects due to the
point-like global monopole spacetime on the matter fields have been
considered for massless scalar \cite{M-L} and fermionic \cite{EVN} fields,
respectively. In order to develop this analysis, the scalar respectively
spinor Green functions in this background have been obtained. The influence
of the non-zero temperature on these polarization effects has been
considered in \cite{C-E} for scalar and fermionic fields. Moreover, the
calculation of quantum effects on massless scalar field in a higher
dimensional global monopole spacetime has also been developed in \cite{E}.
The combined vacuum polarization effects by the non-trivial geometry of a
global monopole and boundary conditions imposed on the matter fields are
investigated as well. In this direction, the total Casimir energy associated
with massive scalar field inside a spherical region in the global monopole
background have been analyzed in Refs. \cite{MKS,EVN1} by using the zeta
function regularization procedure. Scalar Casimir densities induced by
spherical boundaries have been calculated in \cite{A-M,Saha03mon} to higher
dimensional global monopole spacetime by making use of the generalized
Abel-Plana summation formula \cite{Sahmat,Sahrev}. More recently, using also
this formalism, a similar analysis for spinor fields with MIT bag boundary
conditions has been developed in~\cite{Saha-Mello,Beze06}.

Many of treatments of quantum fields around a global monopole deal
mainly with the case of the idealized point-like monopole
geometry. However, the realistic global monopole has a
characteristic core radius determined by the symmetry braking
scale at which the monopole is formed. A simplified model for the
monopole core where the region inside the core is described by the
de Sitter geometry is presented in \cite{Hara90}. The vacuum
polarization effects due to a massless scalar field in the region
outside the core of this model are investigated in Ref.
\cite{Spin05}. In particular, it has been shown that long-range
effects can take place due to the non-trivial core structure. In
the present paper we will analyze the effects of global monopole
core on properties of the quantum vacuum for the general
spherically symmetric static model with a core of finite radius.
The most important quantities characterizing these properties are
the vacuum expectation values of the field square and the
energy-momentum tensor. Though the corresponding operators are
local, due to the global nature of the vacuum, the vacuum
expectation values describe the global properties of the bulk and
carry an important information about the structure of the defect
core. In addition to describing the physical structure of the
quantum field at a given point, the energy-momentum tensor acts as
the source of gravity in the Einstein equations. It therefore
plays an important role in modelling a self-consistent dynamics
involving the gravitational field. As the first step for the
investigation of vacuum densities we evaluate the positive
frequency Wightman function for a massive scalar field with
general curvature coupling parameter. This function gives
comprehensive insight into vacuum fluctuations and determines the
response of a particle detector of the Unruh-DeWitt type moving in
the global monopole bulk. The problem under consideration is also
of separate interest as an example with gravitational and
boundary-induced polarizations of the vacuum, where all
calculations can be performed in a closed form. The corresponding
results specify the conditions under which we can ignore the
details of the interior structure and approximate the effect of
the global monopole by the idealized model.

The paper is organized as follows. In Section \ref{secm:1} we consider the
Wightman function in the exterior of the global monopole for the general
structure of the core assuming that the components of the metric tensor and
their derivatives are continuous at the transition surface between the core
and the exterior. By using this function, in Section \ref{sec:phi2emt} we
investigate the vacuum expectation values of the field square and the
energy-momentum tensor. The Section \ref{sec:thinshell} is devoted to the
generalization of the corresponding results when an additional surface shell
is present on the bounding surface between the core and the exterior. \ As
an illustration of the general results, in Section \ref{sec:flowerpot} we
consider the flower-pot model with the Minkowskian geometry inside the core.
For this model the vacuum expectation values inside the core are
investigated as well. In Section \ref{msec3:conc} we present our concluding
remarks. In Appendix we show that the formulae obtained in the paper for the
core induced parts are also valid in the case when bound states are present.

\section{Wightman function}

\label{secm:1}

We consider a model of $(D+1)$-dimensional global monopole with a core of
radius $a$ in which the spacetime is described by two distinct metric
tensors in the regions outside and inside the core. In the hyperspherical
polar coordinates $(r,\vartheta ,\phi )\equiv (r,\theta _{1},\theta
_{2},\ldots \theta _{n},\phi )$, $n=D-2$, the corresponding line element in
the exterior region $r>a$ has the form
\begin{equation}
ds^{2}=dt^{2}-dr^{2}-\sigma ^{2}r^{2}d\Omega _{D}^{2},  \label{mmetric}
\end{equation}%
where $d\Omega _{D}^{2}$ is the line element on the surface of the unit
sphere in $D$-dimensional Euclidean space, the parameter $\sigma $ is
smaller than unity and is related to the symmetry breaking energy scale in
the theory. The solid angle corresponding to Eq. (\ref{mmetric}) is $\sigma
^{2}S_{D}$ with $S_{D}=2\pi ^{D/2}/\Gamma (D/2)$ being the total area of the
surface of the unit sphere in $D$-dimensional Euclidean space. This leads to
the solid angle deficit $(1-\sigma ^{2})S_{D}$ in the spacetime given by
line element (\ref{mmetric}). It is of interest to note that the effective
metric produced in superfluid $^{3}\mathrm{He-A}$ by a monopole is described
by the three dimensional version of the line element (\ref{mmetric}) with
the negative angle deficit, $\sigma >1$, which corresponds to the negative
mass of the topological object \cite{Volo98}. The quasiparticles in this
model are chiral and massless fermions. We will assume that inside the core
(region $r<a$) the spacetime geometry is regular and is described by the
general static spherically symmetric line element%
\begin{equation}
ds^{2}=e^{2u(r)}dt^{2}-e^{2v(r)}dr^{2}-e^{2w(r)}d\Omega _{D}^{2},
\label{metricinside}
\end{equation}%
where the functions $u(r)$, $v(r)$, $w(r)$ are continuous at the core
boundary:
\begin{equation}
u(a)=v(a)=0,\;w(a)=\ln (\sigma a).  \label{uvbound}
\end{equation}%
Here we assume that there is no surface energy-momentum tensor located at $%
r=a$ and, hence, the derivatives of these functions are continuous as well.
The generalization to the case with an infinitely thin spherical shell at
the boundary of two metrics will be discussed in section \ref{sec:thinshell}%
. Note that by introducing the new radial coordinate $\tilde{r}=e^{w(r)}$
with the core center at $\tilde{r}=0$, the angular part of the line element (%
\ref{metricinside}) is written in the standard Minkowskian form. With this
coordinate, in general, we will obtain non-standard angular part in the
exterior line element (\ref{mmetric}). For the metric corresponding to line
element (\ref{metricinside}) the nonzero components of the Ricci tensor are
given by expressions (no summation over $i$, we adopt the convention of
Birrell and Davies \cite{Birr82} for the curvature tensor)%
\begin{eqnarray}
R_{0}^{0} &=&-e^{-2v}\left[ u^{\prime \prime }+u^{\prime 2}-u^{\prime
}v^{\prime }+(n+1)u^{\prime }w^{\prime }\right] ,  \notag \\
R_{1}^{1} &=&-e^{-2v}\left[ u^{\prime \prime }+u^{\prime 2}-u^{\prime
}v^{\prime }+(n+1)\left( w^{\prime \prime }+w^{\prime 2}-w^{\prime
}v^{\prime }\right) \right] ,  \label{Riccin} \\
R_{i}^{i} &=&-e^{-2v}\left( w^{\prime \prime }+w^{\prime 2}+w^{\prime
}u^{\prime }-w^{\prime }v^{\prime }+nw^{\prime 2}\right) +ne^{-2w},  \notag
\end{eqnarray}%
where the prime means the derivative with respect to the radial coordinate $%
r $ and the indices $i=2,3,\ldots ,D$ correspond to the coordinates $\theta
_{1},\theta _{2},\ldots ,\phi $ respectively. The corresponding Ricci scalar
has the form%
\begin{eqnarray}
R &=&-2e^{-2v}\left[ u^{\prime \prime }+u^{\prime 2}-u^{\prime }v^{\prime
}+n(n+1)w^{\prime 2}/2\right.  \notag \\
&&\left. +(n+1)\left( w^{\prime \prime }+w^{\prime 2}+w^{\prime }u^{\prime
}-w^{\prime }v^{\prime }\right) \right] +n(n+1)e^{-2w}.  \label{Richscin}
\end{eqnarray}%
Note that from the regularity of the interior geometry at the core center
one has the conditions $u(r),v(r)\rightarrow 0$, and $w(r)\sim \ln \tilde{r}$
for $\tilde{r}\rightarrow 0$. In the region outside the core, $r>a$, for the
nonzero components we have the standard expressions (no summation over $i)$:
\begin{equation}
R_{i}^{i}=n\frac{1-\sigma ^{2}}{\sigma ^{2}r^{2}},\quad R=n(n+1)\frac{%
1-\sigma ^{2}}{\sigma ^{2}r^{2}},  \label{mRictens}
\end{equation}%
where $i=2,3,\ldots ,D$. For $n=0$ the spacetime outside the core is flat
and coincides with $D=2$ cosmic string geometry. The influence of the
non-trivial core structure for the cosmic string on a quantum scalar field
has been considered in Refs. \cite{Alle90,Alle96,Beze06b}. In the discussion
below we will assume that $n>0$.

In this paper we are interested in the vacuum polarization effects for a
scalar field with general curvature coupling parameter $\xi $ propagating in
the bulk described above. The corresponding field equation has the form
\begin{equation}
\left( \nabla _{i}\nabla ^{i}+m^{2}+\xi R\right) \varphi =0,
\label{mfieldeq}
\end{equation}%
where $\nabla _{i}$ is the covariant derivative operator associated with
line element (\ref{mmetric}) outside the core and with line element (\ref%
{metricinside}) inside the core. The values of the curvature coupling
parameter $\xi =0$, and $\xi =\xi _{D}$ with $\xi _{D}\equiv (D-1)/4D$
correspond to the most important special cases of minimally and conformally
coupled scalar fields, respectively. As a first stage for the evaluation of
the vacuum expectation values (VEVs) for the field square and the
energy-momentum tensor we consider the positive frequency Wightman function $%
\langle 0|\varphi (x)\varphi (x^{\prime })|0\rangle $, where $|0\rangle $ is
the amplitude for the corresponding vacuum state. This function also
determines the response of the Unruh-DeWitt type particle detector at a
given state of motion (see, for instance, \cite{Birr82}). By expanding the
field operator over eigenfunctions and using the commutation relations one
can see that
\begin{equation}
\langle 0|\varphi (x)\varphi (x^{\prime })|0\rangle =\sum_{\alpha }\varphi
_{\alpha }(x)\varphi _{\alpha }^{\ast }(x^{\prime }),  \label{mfieldmodesum}
\end{equation}%
with $\left\{ \varphi _{\alpha }(x),\varphi _{\alpha }^{\ast }(x^{\prime
})\right\} $ being a complete orthonormalized set of positive and negative
frequency solutions to the field equation. The collective index $\alpha $
can contain both discrete and continuous components. In Eq. (\ref%
{mfieldmodesum}) it is assumed summation over discrete indices and
integration over continuous indices.

Due to the symmetry of the problem under consideration the eigenfunctions
can be presented in the form
\begin{equation}
\varphi _{\alpha }(x)=f_{l}(r)Y(m_{k};\vartheta ,\phi )e^{-i\omega
t},\,\,l=0,1,2,\ldots ,  \label{phialf1}
\end{equation}%
where $m_{k}=(m_{0}\equiv l,m_{1},\ldots ,m_{n})$, and $m_{1},m_{2},\ldots
,m_{n}$ are integers such that
\begin{equation}
0\leq m_{n-1}\leq m_{n-2}\leq \cdots \leq m_{1}\leq l,\quad -m_{n-1}\leq
m_{n}\leq m_{n-1},  \label{mnumbvalues}
\end{equation}%
and $Y(m_{k};\vartheta ,\phi )$ is the hyperspherical harmonic of degree $l$
\cite{Erdelyi}. The equation for the radial function is obtained from the
field equation (\ref{mfieldeq}) and has the form%
\begin{equation}
f_{l}^{\prime \prime }(r)+\left[ u^{\prime }-v^{\prime }+(D-1)w^{\prime }%
\right] f_{l}^{\prime }(r)+e^{2v}\left[ e^{-2u}\omega ^{2}-m^{2}-\xi
R-l(l+n)e^{-2w}\right] f_{l}(r)=0.  \label{fleq}
\end{equation}%
In the region $r>a$ described by the line element (\ref{mmetric}), the
linearly independent solutions to this equation are $r^{-n/2}J_{\nu
_{l}}(\lambda r)$ and $r^{-n/2}Y_{\nu _{l}}(\lambda r)$ with $\lambda =\sqrt{%
\omega ^{2}-m^{2}}$, where $J_{\nu _{l}}(x)$ and $Y_{\nu _{l}}(x)$ are the
Bessel and Neumann functions with the order
\begin{equation}
\nu _{l}=\frac{1}{\sigma }\left[ \left( l+\frac{n}{2}\right) ^{2}+(1-\sigma
^{2})n(n+1)\left( \xi -\xi _{D-1}\right) \right] ^{\frac{1}{2}}.
\label{nuel}
\end{equation}%
In the following consideration we will assume that $\nu _{l}^{2}$ is
non-negative. This corresponds to the restriction on the values of the
curvature coupling parameter for $n>0$, given by the condition
\begin{equation}
(1-\sigma ^{2})\xi \geqslant -\frac{n\sigma ^{2}}{4(n+1)}.  \label{condposnu}
\end{equation}%
This condition is satisfied by the minimally coupled field for all values $%
\sigma $ and by the conformally coupled field for $\sigma \leqslant D-1$.
The solution of the radial equation (\ref{fleq}) in the region $r<a$ regular
at the origin we will denote by $R_{l}(r,\lambda )$. From Eq. (\ref{fleq})
it follows that near the core center this solution behaves as $\tilde{r}^{l}$%
. Note that the parameter $\lambda $ enters in the radial equation in the
form $\lambda ^{2}$. As a result the regular solution can be chosen in such
a way that $R_{l}(r,-\lambda )=\mathrm{const}\cdot R_{l}(r,\lambda )$. Now
for the radial part of the eigenfunctions one has%
\begin{equation}
f_{l}(r)=\left\{
\begin{array}{ll}
R_{l}(r,\lambda ) & \mathrm{for}\;r<a \\
r^{-n/2}\left[ A_{l}J_{\nu _{l}}(\lambda r)+B_{l}Y_{\nu _{l}}(\lambda r)%
\right] & \mathrm{for}\;r>a%
\end{array}%
,\right.  \label{fl}
\end{equation}%
where the coefficients $A_{l}$ and $B_{l}$ are determined by the conditions
of continuity of the radial function and its derivative at $r=a$. From these
conditions we find the following expressions for these coefficients%
\begin{equation}
A_{l}=\frac{\pi }{2}a^{n/2}R_{l}(a,\lambda )\bar{Y}_{\nu _{l}}(\lambda
a),\;B_{l}=-\frac{\pi }{2}a^{n/2}R_{l}(a,\lambda )\bar{J}_{\nu _{l}}(\lambda
a).  \label{AlBl}
\end{equation}%
Here and in what follows for a cylinder function $F(z)$ we use the notation
\begin{equation}
\bar{F}(z)\equiv zF^{\prime }(z)-\left[ \frac{n}{2}+a\frac{R_{l}^{\prime
}(a,z/a)}{R_{l}(a,z/a)}\right] F(z),  \label{barnot}
\end{equation}%
where $R_{l}^{\prime }(a,\lambda )=\partial R_{l}(r,\lambda )/\partial
r|_{r=a}$. Note that due to our choice of the function $R_{l}(r,\lambda )$,
the logarithmic derivative in formula (\ref{barnot}) is an even function on $%
z$. Hence, in the region $r>a$ the radial part of the eigenfunctions has the
form%
\begin{equation}
f_{l}(r)=\frac{\pi a^{n/2}}{2r^{n/2}}R_{l}(a,\lambda )g_{\nu _{l}}(\lambda
a,\lambda r),  \label{fl2}
\end{equation}%
where the notation
\begin{equation}
g_{\nu _{l}}(\lambda a,\lambda r)=J_{\nu _{l}}(\lambda r)\bar{Y}_{\nu
_{l}}(\lambda a)-\bar{J}_{\nu _{l}}(\lambda a)Y_{\nu _{l}}(\lambda r).
\label{gnot}
\end{equation}%
is introduced.

For the eigenfunctions we have the following orthonormalization condition%
\begin{equation}
\int dV\sqrt{-g}g^{00}\varphi _{\alpha }(x)\varphi _{\alpha ^{\prime
}}^{\ast }(x)=\frac{\delta _{\alpha \alpha ^{\prime }}}{2\omega },
\label{ortnorm}
\end{equation}%
where $\delta _{\alpha \alpha ^{\prime }}$ is understood as the Kronecker
symbol for discrete indices and as the Dirac delta function for continuous
ones. Substituting eigenfunctions (\ref{phialf1}), and using the relation
\begin{equation}
\int d\Omega \,\left\vert Y(m_{k};\vartheta ,\phi )\right\vert ^{2}=N(m_{k})
\label{harmint}
\end{equation}%
(the explicit form for $N(m_{k})$ is given in \cite{Erdelyi} and will not be
necessary for the following consideration in this paper), the normalization
condition is written in terms of the radial eigenfunctions
\begin{equation}
\int_{r_{0}}^{\infty }dr\,\sqrt{-g_{r}}g^{00}f_{l}(r,\lambda
)f_{l}(r,\lambda ^{\prime })=\frac{\delta (\lambda -\lambda ^{\prime })}{%
2\omega N(m_{k})},  \label{normfl}
\end{equation}%
where $r_{0}$ is the value of the radial coordinate $r$ corresponding to the
origin and $g_{r}$ is the radial part of the determinant $g$. Note that in
general $r_{0}\neq 0$ (see, for instance, the special case of the flower-pot
model in section \ref{sec:flowerpot}). As the integral on the left is
divergent for $\lambda ^{\prime }=\lambda $, the main contribution in the
coincidence limit comes from large values $r$. \ By using the expression (%
\ref{fl2}) for the radial part in the region $r>a$ and replacing the Bessel
and Neumann functions by the leading terms of their asymptotic expansions
for large values of the argument, it can be seen that from (\ref{normfl})
the following result is obtained:%
\begin{equation}
a^{n}R_{l}^{2}(a,\lambda )=\frac{2\sigma ^{1-D}\lambda }{\pi ^{2}\omega
N(m_{k})\left[ \bar{J}_{\nu _{l}}^{2}(\lambda a)+\bar{Y}_{\nu
_{l}}^{2}(\lambda a)\right] }.  \label{normcoefRl}
\end{equation}

Having the normalized eigenfunctions, now we turn to the evaluation of the
Wightman function by using the mode sum formula (\ref{mfieldmodesum}).
Substituting eigenfunctions (\ref{fl2}) and using the addition formula for
the hyperspherical harmonics \cite{Erdelyi}
\begin{equation}
\sum_{m_{k}}\frac{Y(m_{k};\vartheta ,\phi )}{N(m_{k})}Y^{\ast
}(m_{k};\vartheta ^{\prime },\phi ^{\prime })=\frac{2l+n}{nS_{D}}%
C_{l}^{n/2}(\cos \theta ),  \label{adtheorem}
\end{equation}%
for the Wightman function in the region outside the monopole's core one
obtains
\begin{eqnarray}
\langle 0|\varphi (x)\varphi (x^{\prime })|0\rangle &=&\frac{\sigma ^{1-D}}{%
2nS_{D}}\sum_{l=0}^{\infty }\frac{2l+n}{(rr^{\prime })^{n/2}}%
C_{l}^{n/2}(\cos \theta )  \notag \\
&\times &\int_{0}^{\infty }\frac{\lambda d\lambda }{\sqrt{\lambda ^{2}+m^{2}}%
}\frac{g_{\nu _{l}}(\lambda a,\lambda r)g_{\nu _{l}}(\lambda a,\lambda
r^{\prime })}{\bar{J}_{\nu _{l}}^{2}(\lambda a)+\bar{Y}_{\nu
_{l}}^{2}(\lambda a)}e^{i\sqrt{\lambda ^{2}+m^{2}}(t^{\prime }-t)}.
\label{unregWightout}
\end{eqnarray}%
In formula (\ref{adtheorem}), $S_{D}=2\pi ^{D/2}/\Gamma (D/2)$ is the total
area of the surface of the unit sphere in $D$-dimensional space, $%
C_{p}^{q}(x)$ is the Gegenbauer or ultraspherical polynomial of degree $p$
and order $q$, and $\theta $ is the angle between directions $(\vartheta
,\phi )$ and $(\vartheta ^{\prime },\phi ^{\prime })$. Let us denote by $%
\langle 0_{\mathrm{m}}|\varphi (x)\varphi (x^{\prime })|0_{\mathrm{m}%
}\rangle $ the positive frequency Wightman function for the geometry of the
idealized point-like global monopole described by the line element (\ref%
{mmetric}) for all values of the radial coordinate. This function can be
presented in the form \cite{A-M}
\begin{eqnarray}
\langle 0_{\mathrm{m}}|\varphi (x)\varphi (x^{\prime })|0_{\mathrm{m}%
}\rangle &=&\frac{\sigma ^{1-D}}{2nS_{D}}\sum_{l=0}^{\infty }\frac{2l+n}{%
(rr^{\prime })^{n/2}}C_{l}^{n/2}(\cos \theta )  \notag \\
&&\times \int_{0}^{\infty }d\lambda \,\frac{\lambda e^{i\sqrt{\lambda
^{2}+m^{2}}(t^{\prime }-t)}}{\sqrt{\lambda ^{2}+m^{2}}}J_{\nu _{l}}(\lambda
r)J_{\nu _{l}}(\lambda r^{\prime }).  \label{Mink}
\end{eqnarray}%
In order to investigate the part induced by the non-trivial core structure,
we consider the difference
\begin{equation}
\langle \varphi (x)\varphi (x^{\prime })\rangle _{c}=\langle 0|\varphi
(x)\varphi (x^{\prime })|0\rangle -\langle 0_{\mathrm{m}}|\varphi (x)\varphi
(x^{\prime })|0_{\mathrm{m}}\rangle .  \label{coreWF}
\end{equation}%
Using formulae (\ref{unregWightout}), (\ref{Mink}) and the relation
\begin{equation}
\frac{g_{\nu }(\lambda a,\lambda r)g_{\nu }(\lambda a,\lambda r^{\prime })}{%
\bar{J}_{\nu }^{2}(\lambda a)+\bar{Y}_{\nu }^{2}(\lambda a)}-J_{\nu
}(\lambda r)J_{\nu }(\lambda r^{\prime })=-\frac{1}{2}\sum_{s=1}^{2}\frac{%
\bar{J}_{\nu }(\lambda a)}{\bar{H}_{\nu }^{(s)}(\lambda a)}H_{\nu
}^{(s)}(\lambda r)H_{\nu }^{(s)}(\lambda r^{\prime }),  \label{relab}
\end{equation}%
with $H_{\nu }^{(s)}(x)$, $s=1,2$ being the Hankel functions, the core
induced part in the Wightman function is presented in the form
\begin{eqnarray}
\langle \varphi (x)\varphi (x^{\prime })\rangle _{c} &=&-\frac{\sigma ^{1-D}%
}{4nS_{D}}\sum_{l=0}^{\infty }\frac{2l+n}{(rr^{\prime })^{n/2}}%
C_{l}^{n/2}(\cos \theta )\sum_{s=1}^{2}\int_{0}^{\infty }d\lambda \,\lambda
\notag \\
&&\times \frac{e^{i\sqrt{\lambda ^{2}+m^{2}}(t^{\prime }-t)}}{\sqrt{\lambda
^{2}+m^{2}}}\frac{\bar{J}_{\nu _{l}}(\lambda a)}{\bar{H}_{\nu
_{l}}^{(s)}(\lambda a)}H_{\nu _{l}}^{(s)}(\lambda r)H_{\nu
_{l}}^{(s)}(\lambda r^{\prime }).  \label{extdif}
\end{eqnarray}%
Now we rotate the integration contour in the complex plane $\lambda $ by the
angle $\pi /2$ for $s=1$ and by the angle $-\pi /2$ for $s=2$. By using the
property that the logarithmic derivative of the function $R_{l}(r,\lambda )$
in formula (\ref{barnot}) is an even function on $z$, we can see that the
integrals over the segments $(0,im)$ and $(0,-im)$ of the imaginary axis
cancel out. As a result, after introducing the modified Bessel functions,
the core induced part can be presented in the form
\begin{eqnarray}
\langle \varphi (x)\varphi (x^{\prime })\rangle _{c} &=&-\frac{\sigma ^{1-D}%
}{\pi nS_{D}}\sum_{l=0}^{\infty }\frac{2l+n}{(rr^{\prime })^{n/2}}%
C_{l}^{n/2}(\cos \theta )\int_{m}^{\infty }dz\,z\frac{\tilde{I}_{\nu
_{l}}(za)}{\tilde{K}_{\nu _{l}}(za)}  \notag \\
&&\times \frac{K_{\nu _{l}}(zr)K_{\nu _{l}}(zr^{\prime })}{\sqrt{z^{2}-m^{2}}%
}\cosh \!\left[ \sqrt{z^{2}-m^{2}}(t^{\prime }-t)\right] .
\label{regWightout}
\end{eqnarray}%
Here and below the tilted notation for the modified Bessel functions is
defined as%
\begin{equation}
\tilde{F}(z)\equiv zF^{\prime }(z)-\mathcal{R}_{l}(a,z)F(z),
\label{Barrednotmod}
\end{equation}%
with%
\begin{equation}
\mathcal{R}_{l}(a,z)=\frac{n}{2}+a\frac{R_{l}^{\prime }(a,ze^{\pi i/2}/a)}{%
R_{l}(a,ze^{\pi i/2}/a)}.  \label{Rlcal}
\end{equation}%
The VEVs in the bulk of the idealized point-like global monopole are
well-investigated in literature (see, for instance, \cite{M-L}-\cite%
{Saha03mon} and references therein) and in the discussion below we will be
mainly concerned with the part induced by the non-trivial core structure. As
we see from (\ref{regWightout}), all information about the inner structure
of the global monopole is contained in the logarithmic derivative of the
interior radial function in formula (\ref{Rlcal}). In deriving formula (\ref%
{regWightout}) we have assumed that there are no bound states for which $%
\lambda $ is purely imaginary. In appendix we show that this formula is also
valid in the case when bound states are present.

\section{Vacuum expectation values outside the monopole core}

\label{sec:phi2emt}

The VEV of the field square is obtained by computing the Wightman function
in the coincidence limit $x^{\prime }\rightarrow x$. In this limit
expression (\ref{unregWightout}) gives a divergent result and some
renormalization procedure is needed. Outside the monopole core the local
geometry is the same as that for a point-like global monopole. Hence, in the
region $r>a$ the renormalization procedure for the local characteristics of
the vacuum, such as the field square and the energy-momentum tensor, is the
same as for the point-like global monopole geometry. This procedure is
discussed in a number of papers (see \cite{M-L}--\cite{E}). For the
renormalization we must subtract the corresponding DeWitt--Schwinger
expansion involving the terms up to order $D$. For a massless field the
renormalized value of the field square has the structure $\langle \varphi
^{2}\rangle _{\mathrm{m,ren}}=\left[ A+B\ln (\mu r)\right] /r^{D-1}$, where
the coefficients $A$ and $B$ are functions on the parameters $\sigma $ and $%
\xi $ only and the arbitrary mass scale $\mu $ corresponds to the ambiguity
in the renormalization procedure. For a spacetime of odd dimension $B=0$ and
this ambiguity is absent. In general, it is not possible to obtain closed
expression for the coefficients $A$ and $B$. For small values $1-\sigma ^{2}$
approximate expressions are derived in Ref. \cite{E} for $D=4$ and $D=5$. In
this paper our main interest are the parts in the VEVs induced by the
non-trivial core structure and below we will concentrate on these quantities.

By using the formula for the Wightman function from the previous section,
the VEV of the field square in the exterior region is presented in the form%
\begin{equation*}
\langle \varphi ^{2}\rangle _{\mathrm{ren}}=\langle \varphi ^{2}\rangle _{%
\mathrm{m,ren}}+\langle \varphi ^{2}\rangle _{c}
\end{equation*}%
Taking into account the relation
\begin{equation}
C_{l}^{n/2}(1)=\frac{\Gamma (l+n)}{\Gamma (n)l!},  \label{Cl1}
\end{equation}%
for the part induced by the core we find
\begin{equation}
\langle \varphi ^{2}\rangle _{c}=-\frac{\sigma ^{1-D}}{\pi r^{n}S_{D}}%
\sum_{l=0}^{\infty }D_{l}\int_{m}^{\infty }dz\ z\frac{\tilde{I}_{\nu
_{l}}(za)}{\tilde{K}_{\nu _{l}}(za)}\frac{K_{\nu _{l}}^{2}(zr)}{\sqrt{%
z^{2}-m^{2}}}.  \label{regsquareout}
\end{equation}%
Here the factor
\begin{equation}
D_{l}=(2l+D-2)\frac{\Gamma (l+D-2)}{\Gamma (D-1)\,l!}  \label{Dlang}
\end{equation}%
is the degeneracy of each angular mode with given $l$. For a fixed $l$ and
large $z$ the integrand contains the exponential factor $e^{2z(a-r)}$ and
the integral converges when $r>a$. For large values $l$, introducing a new
integration variable $y=z/\nu _{l}$ in the integral of Eq. (\ref%
{regsquareout}) and using the uniform asymptotic expansions for the modified
Bessel functions \cite{Abra64}, it can be seen that the both integral and
sum are convergent for $r>a$ and diverge at $r=a$. For the points near the
sphere the part (\ref{regsquareout}) behaves as $1/(r-a)^{\beta _{1}}$,
where $\beta _{1}$ is an integer which depends on the specific model of the
core. For this parameter one has $\beta _{1}\leqslant D-1$. The exception is
the case of the core model for which the leading term in the uniform
asymptotic expansion of the function $\tilde{K}_{\nu _{l}}(za)$ for large
values $l$ vanishes. The latter takes place for the interior radial function
with the asymptotic behavior $\mathcal{R}_{l}(a,lz/\sigma )\sim -(l/\sigma )%
\sqrt{1+z^{2}}$ for large $l$. For the case of a massless scalar the
asymptotic behavior of the part (\ref{regsquareout}) at large distances from
the sphere can be obtained by introducing a new integration variable $y=zr$
and expanding the integrand in terms of $a/r$. The leading contribution for
the summand with a given $l$ has an order $(a/r)^{2\nu _{l}+D-1}$ [assuming
that $\nu _{l}\neq 0$ and $\mathcal{R}_{l}(a,0)\neq \pm \nu _{l}$] and the
main contribution comes from the $l=0$ term. Now comparing this with the
part $\langle \varphi ^{2}\rangle _{\mathrm{m,ren}}$, we see that for $\nu
_{0}>0$ the VEV of the field square at large distances from the core is
dominated by the part corresponding to the geometry of the point-like global
monopole. For the case $\nu _{0}=0$ the ratio $\langle \varphi ^{2}\rangle
_{c}/\langle \varphi ^{2}\rangle _{\mathrm{m,ren}}$ decays logarithmically
and long-range effects of the monopole core appear similar to those for the
geometry of a cosmic string \cite{Alle90,Alle96} (see also the discussion in
Ref. \cite{Spin05} for the model with de Sitter spacetime inside the core).
This case is realized by special values of the parameters satisfying the
condition $(1/\sigma ^{2}-1)\xi =-\xi _{D-1}$. For a massive field assuming
that $mr\gg 1$, the main contribution into the integral over $z$ in Eq. (\ref%
{regsquareout}) comes from the lower limit and to the leading order one has
\begin{equation}
\langle \varphi ^{2}\rangle _{c}=-\frac{\sqrt{\pi }\sigma ^{1-D}e^{-2mr}}{%
4r^{n+1}S_{D}\sqrt{mr}}\sum_{l=0}^{\infty }D_{l}\frac{\tilde{I}_{\nu
_{l}}(ma)}{\tilde{K}_{\nu _{l}}(ma)},  \label{phi2massivlarger}
\end{equation}%
with the exponentially suppressed VEV.

Consider the limit $\sigma \ll 1$ for a fixed value $r$. In accordance with
Eq. (\ref{mRictens}) this corresponds to large values of the scalar
curvature and, hence, to strong gravitational fields. To satisfy condition (%
\ref{condposnu}) we will assume that $\xi \geqslant 0$. For $\xi >0$, from
Eq. (\ref{nuel}) one has $\nu _{l}\gg 1$, and after introducing in Eq. (\ref%
{regsquareout}) a new integration variable $y=z/\nu _{l}$, we can replace
the modified Bessel function by their uniform asymptotic expansions for
large values of the order. The main contribution to the sum over $l$ comes
from the summand with $l=0$, and the core induced VEV $\langle \varphi
^{2}\rangle _{c}$ is suppressed by the factor $\exp \left[ -(2/\sigma )\sqrt{%
n(n+1)\xi }\ln (r/a)\right] $. For $\xi =0$ and $\sigma \ll 1$ for the terms
with $l\neq 0$ one has $\nu _{l}\gg 1$ and the corresponding contribution is
again exponentially small. For the summand with $l=0$ to the leading order
over $\sigma $ we have $\nu _{l}=n/2$ and $\langle \varphi ^{2}\rangle
_{c}\sim 1/\sigma ^{D-1}$. Hence, we conclude that in the limit of strong
gravitational fields the behavior of the VEV $\langle \varphi ^{2}\rangle
_{c}$ is completely different for minimally and non-minimally coupled
scalars.

Now we turn to the investigation of the VEV of the energy-momentum tensor in
the region $r>a$. Having the Wightman function and the VEV\ for the field
square, these VEVs are evaluated on the base of the formula
\begin{equation}
\langle 0|T_{ik}|0\rangle =\lim_{x^{\prime }\rightarrow x}\partial
_{i}\partial _{k}^{\prime }\langle 0|\varphi (x)\varphi (x^{\prime
})|0\rangle +\left[ \left( \xi -\frac{1}{4}\right) g_{ik}\nabla _{l}\nabla
^{l}-\xi \nabla _{i}\nabla _{k}-\xi R_{ik}\right] \langle 0|\varphi
^{2}(x)|0\rangle .  \label{mvevEMT}
\end{equation}%
Similar to the Wightman function, the components of the vacuum
energy-momentum tensor can be presented in the decomposed form%
\begin{equation}
\langle 0|T_{ik}|0\rangle =\langle 0_{\mathrm{m}}|T_{ik}|0_{\mathrm{m}%
}\rangle +\langle T_{ik}\rangle _{c},  \label{Tikextdecomp}
\end{equation}%
where $\langle 0_{\mathrm{m}}|T_{ik}|0_{\mathrm{m}}\rangle $ is the vacuum
energy-momentum tensor for the geometry of a point-like global monopole and
the part $\langle T_{ik}\rangle _{c}$ is induced by the core. In accordance
with the problem symmetry both these tensors are diagonal. For massless
fields the VEV of the energy-momentum tensor for the point-like global
monopole geometry is investigated in Refs. \cite{M-L}--\cite{E}. The
corresponding renormalized components have the structure similar to that
given for the field square:%
\begin{equation}
\langle T_{ik}\rangle _{\mathrm{m,ren}}=\frac{1}{r^{D+1}}\left[
q_{ik}^{(1)}+q_{ik}^{(2)}\ln (\mu r)\right] ,  \label{Tikmren}
\end{equation}%
where the coefficients $q_{ik}^{(1)}$, $q_{ik}^{(2)}$ depend only on the
parameters $\sigma $ and $\xi $, and $q_{ik}^{(2)}=0$ for $D$ being an even
number. Substituting the expressions of the Wightman function and the VEV of
the field square into formula (\ref{mvevEMT}), for the part of the
energy-momentum tensor induced by the non-trivial core structure one obtains
(no summation over $i$)
\begin{equation}
\langle T_{i}^{k}\rangle _{c}=-\frac{\sigma ^{1-D}\delta _{i}^{k}}{2\pi
r^{n}S_{D}}\sum_{l=0}^{\infty }D_{l}\int_{m}^{\infty }dz\,z^{3}\frac{\tilde{I%
}_{\nu _{l}}(za)}{\tilde{K}_{\nu _{l}}(za)}\frac{F_{\nu _{l}}^{(i)}\left[
K_{\nu _{l}}(zr)\right] }{\sqrt{z^{2}-m^{2}}},\quad r>a,  \label{q1in}
\end{equation}%
where for a given function $f(y)$ the notations
\begin{eqnarray}
F_{\nu _{l}}^{(0)}\left[ f(y)\right] &=&(1-4\xi )\left[ f^{^{\prime }2}(y)-%
\frac{n}{y}f(y)f^{\prime }(y)+\left( \frac{\nu _{l}^{2}}{y^{2}}-\frac{1+4\xi
-2(mr/y)^{2}}{1-4\xi }\right) f^{2}(y)\right] ,  \label{Fineps} \\
F_{\nu _{l}}^{(1)}\left[ f(y)\right] &=&f^{^{\prime }2}(y)+\frac{\tilde{\xi}%
}{y}f(y)f^{\prime }(y)-\left( 1+\frac{\nu _{l}^{2}+\tilde{\xi}n/2}{y^{2}}%
\right) f^{2}(y),  \label{Finperad} \\
F_{\nu _{l}}^{(i)}\left[ f(y)\right] &=&(4\xi -1)f^{^{\prime }2}(y)-\frac{%
\tilde{\xi}}{y}f(y)f^{\prime }(y)+\left[ 4\xi -1+\frac{\nu _{l}^{2}(1+\tilde{%
\xi})+\tilde{\xi}n/2}{(n+1)y^{2}}\right] f^{2}(y),  \label{Finpeaz}
\end{eqnarray}%
are introduced with $\tilde{\xi}=4(n+1)\xi -n$ and in Eq. (\ref{Finpeaz}) $%
i=2,3,\ldots ,D$. It can be seen that components (\ref{q1in}) satisfy the
continuity equation $\nabla _{k}\langle T_{i}^{k}\rangle _{c}=0$, which for
the geometry under consideration takes the form%
\begin{equation}
r\frac{d}{dr}\langle T_{1}^{1}\rangle _{c}+(D-1)\left( \langle
T_{1}^{1}\rangle _{c}-\langle T_{2}^{2}\rangle _{c}\right) =0.
\label{conteq1}
\end{equation}

The core induced part $\langle T_{i}^{k}\rangle _{c}$ are finite everywhere
outside the core, $r>a$, and diverge on the core boundary. Near this
boundary the main contribution comes from large values $l$ and to find the
corresponding asymptotic behavior we can use the uniform asymptotic
expansions for the modified Bessel functions. To the leading order one finds
$\langle T_{i}^{k}\rangle _{c}\sim 1/(r-a)^{\beta _{2}}$ for the energy
density and the azimuthal stress and
\begin{equation}
\langle T_{1}^{1}\rangle _{c}\approx -\frac{D-1}{\beta _{2}-1}(r/a-1)\langle
T_{2}^{2}\rangle _{c},  \label{T11T22}
\end{equation}
with $\beta _{2}\leqslant D+1$. An exception is the special case of the core
model for which the leading term in the uniform asymptotic expansion for the
function $\tilde{K}_{\nu _{l}}(za)$ vanishes. For large distances from the
core boundary, $r\gg a$, and for a massless scalar field the main
contribution into the VEV $\langle T_{i}^{k}\rangle _{c}$ comes from the $%
l=0 $ summand. Under the assumptions $\nu _{0}\neq 0$ and $\mathcal{R}%
_{0}(a,0)\neq \pm \nu _{0}$, the leading term of the corresponding
asymptotic expansion behaves like $\langle T_{i}^{k}\rangle _{c}\sim
(a/r)^{2\nu _{0}+D+1}$. For a massive scalar field under the condition $%
mr\gg 1$, the main contribution into the integral over $z$ in Eq. (\ref{q1in}%
) comes from the lower limit and by using the asymptotic formulae for the
function $K_{\nu _{l}}(zr)$ for large values of the argument, to the leading
order one finds%
\begin{equation}
\langle T_{0}^{0}\rangle _{c}\approx -\langle T_{2}^{2}\rangle _{c}\approx
(4\xi -1)\frac{\sqrt{\pi }m^{3/2}e^{-2mr}}{4r^{D-1/2}S_{D}\sigma ^{D-1}}%
\sum_{l=0}^{\infty }D_{l}\frac{\tilde{I}_{\nu _{l}}(ma)}{\tilde{K}_{\nu
_{l}}(ma)},  \label{Tikcmassivefar}
\end{equation}%
and the radial stress is suppressed by an additional factor $1/mr$.

Now let us consider the VEV of the energy-momentum tensor in the limit $%
\sigma \ll 1$ for a fixed $r>a$. For $\xi >0$ by the calculations similar to
those given above for the field square, one finds that the core induced VEV
are suppressed by the factor $\exp \left[ -(2/\sigma )\sqrt{n(n+1)\xi }\ln
(r/a)\right] $ and the vacuum stresses are strongly anisotropic: $\langle
T_{1}^{1}\rangle _{c}/\langle T_{2}^{2}\rangle _{c}\sim \sigma $. For a
minimally coupled scalar, $\xi =0$, the leading term of the asymptotic
expansion over $\sigma $ comes from the $l=0$ summand in Eq. (\ref{q1in})
with $\nu _{l}=n/2$. This term behaves as $\sigma ^{1-D}$.

\section{Core with an infinitely thin shell}

\label{sec:thinshell}

The results considered in the previous section can be generalized to the
models where an additional infinitely thin spherical shell located at $r=a$
is present with the surface energy-momentum tensor $\tau _{i}^{k}$. We
denote by $n^{i}$ the normal to the shell normalized by the condition $%
n_{i}n^{i}=-1$, assuming that it points into the bulk on both sides. From
the Israel matching conditions one has%
\begin{equation}
\left\{ K_{ik}-Kh_{ik}\right\} =8\pi G\tau _{ik},  \label{matchcond}
\end{equation}%
where the curly brackets denote summation over each side of the shell, $%
h_{ik}=g_{ik}+n_{i}n_{k}$ is the induced metric on the shell, $%
K_{ik}=h_{i}^{r}h_{k}^{s}\nabla _{r}n_{s}$ its extrinsic curvature and $%
K=K_{i}^{i}$. For the region $r\leqslant a$ one has $n_{i}=\delta
_{i}^{1}e^{v(r)}$ and the non-zero components of the extrinsic curvature are
given by the formulae%
\begin{equation}
K_{0}^{0}=-u^{\prime }(r)e^{-v(r)},\;K_{i}^{j}=-\delta _{i}^{j}w^{\prime
}(r)e^{-v(r)},\;i=2,3,\ldots ,\;r=a-.  \label{exttensor}
\end{equation}%
The corresponding expressions for the region $r\geqslant a$ are obtained by
taking $u(r)=v(r)=0$, $w(r)=\ln (\sigma r)$ and changing the signs for the
components of the extrinsic curvature tensor. Now from the matching
conditions (\ref{matchcond}) we find (no summation over $i$)%
\begin{eqnarray}
u^{\prime }(a-) &=&8\pi G\left[ \tau _{i}^{i}-\frac{D-2}{D-1}\tau _{0}^{0}%
\right] ,\;i=2,3,\ldots ,  \label{matchcond1} \\
w^{\prime }(a-) &=&\frac{1}{a}+\frac{8\pi G}{D-1}\tau _{0}^{0}
\label{matchcond2}
\end{eqnarray}%
where $f^{\prime }(a-)$ is understood in the sense $\lim_{r\rightarrow
a-0}f^{\prime }(r)$. The discontinuity of the functions $u^{\prime }(r)$ and
$w^{\prime }(r)$ at $r=a$ leads to the delta function term%
\begin{equation}
2\left[ u^{\prime }(a-)+(D-1)\left( w^{\prime }(a-)-1/a\right) \right]
\delta (r-a)  \label{deltain Ricci}
\end{equation}%
in the Ricci scalar and, hence, in the equation (\ref{fleq}) for the radial
eigenfunctions. Note that the expression in the square brackets is related
to the surface energy-momentum tensor by the formula%
\begin{equation}
u^{\prime }(a-)+(D-1)\left( w^{\prime }(a-)-1/a\right) =\frac{8\pi G}{D-1}%
\tau ,  \label{Tracesurf}
\end{equation}%
where $\tau $ is the trace of the surface energy-momentum tensor.

Due to the delta function term in the equation for the radial
eigenfunctions, these functions have a discontinuity in their slope at $r=a$%
. The corresponding jump condition is obtained by integrating the equation (%
\ref{fleq}) through the point $r=a$:%
\begin{equation}
f_{l}^{\prime }(a+)-f_{l}^{\prime }(a-)=\frac{16\pi G\xi }{D-1}\tau f_{l}(a).
\label{flderjump}
\end{equation}%
Now the coefficients in the formulae (\ref{fl}) for the eigenfunctions are
determined by the continuity condition for the radial eigenfunctions and by
the jump condition for their radial derivative. It can be seen that the
corresponding eigenfunctions are given by the same formulae (\ref{fl2}) and (%
\ref{normcoefRl}) with the new barred notation%
\begin{equation}
\bar{F}(z)\equiv zF^{\prime }(z)-\left[ \frac{n}{2}+\frac{16\pi G\xi }{D-1}%
a\tau +a\frac{R_{l}^{\prime }(a,\lambda )}{R_{l}(a,\lambda )}\right] F(z).
\label{barrednew}
\end{equation}%
Consequently the parts in the Wightman function, in the VEVs of the field
square and the energy-momentum tensor induced by the core of the finite
thickness, are given by formula (\ref{regWightout}), (\ref{regsquareout})
and (\ref{q1in}), where the tilted notation is defined by Eq. (\ref%
{Barrednotmod}) with the function%
\begin{equation}
\mathcal{R}_{l}(a,z)=\frac{n}{2}+\frac{16\pi G\xi }{D-1}a\tau +a\frac{%
R_{l}^{\prime }(a,ze^{\pi i/2}/a)}{R_{l}(a,ze^{\pi i/2}/a)}.
\label{newRlcal}
\end{equation}%
The trace of the surface energy-momentum tensor in this expression is
related to the components of the metric tensor inside the core by formula (%
\ref{Tracesurf}).

\section{Flower-pot model for global monopole}

\label{sec:flowerpot}

As an application of the general results given above let us consider a
simple example of the core model assuming that the spacetime inside the core
is flat. The corresponding model for the cosmic string core was considered
in Refs. \cite{Alle90,Alle96,Beze06b} and following these papers we will
refer to this model as flower-pot model. Taking $u(r)=v(r)=0$ from the zero
curvature condition one finds $e^{w(r)}=r+\mathrm{const}$. The value of the
constant here is found from the continuity condition for the function $w(r)$
at the boundary which gives $\mathrm{const}=(\sigma -1)a$. Hence, the
interior line element has the form%
\begin{equation}
ds^{2}=dt^{2}-dr^{2}-\left[ r+(\sigma -1)a\right] ^{2}d\Omega _{D}^{2}.
\label{intflow}
\end{equation}%
In terms of the radial coordinate $r$ the origin is located at $r=(1-\sigma
)a$. From the matching conditions (\ref{matchcond1}), (\ref{matchcond2}) we
find the corresponding surface energy-momentum tensor with the non-zero
components%
\begin{equation}
\tau _{0}^{0}=\left( \frac{1}{\sigma }-1\right) \frac{D-1}{8\pi Ga},\;\tau
_{i}^{k}=\frac{D-2}{D-1}\tau _{0}^{0}\delta _{i}^{k},\;i=2,3,\ldots .
\label{surfemtflow}
\end{equation}%
The corresponding surface energy density is positive for the global monopole
with $\sigma <1$. After this brief review, let us analyze for this model the
influence of the monopole's core on the vacuum polarization effects. We will
consider the exterior and interior regions separately.

\subsection{Exterior region}

\label{subsec:flpotExt}

\begin{figure}[tbph]
\begin{center}
\epsfig{figure=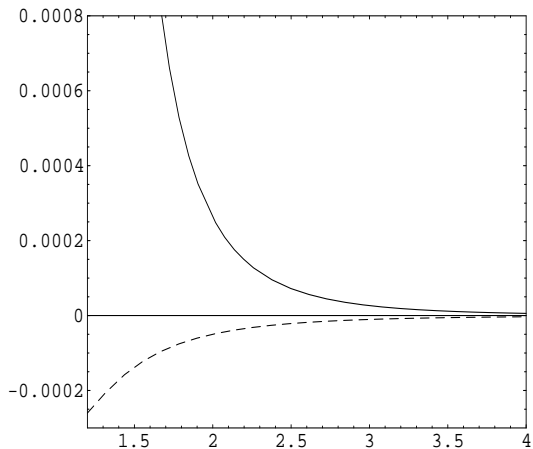,width=7.cm,height=6.cm}
\end{center}
\caption{The expectation value $a^{D-1}\langle \protect\varphi ^{2}\rangle
_{c}$ induced by the non-trivial core structure in the region outside the
core for $D=3$ massless scalar field as a function of $r/a$ in flower-pot
model with $\protect\sigma =0.5$. The full/dashed curves correspond to
minimally/conformally coupled scalars.}
\label{fig1}
\end{figure}

In the region inside the core the radial eigenfunctions regular at the
origin are the functions:%
\begin{equation}
R_{l}(r,\lambda )=C_{l}\frac{J_{l+n/2}(\lambda \tilde{r})}{\tilde{r}^{n/2}},
\label{Rlflow}
\end{equation}%
where $\tilde{r}=r+(\sigma -1)a$ is the standard Minkowskian radial
coordinate, $0\leqslant \tilde{r}\leqslant \sigma a$. In appendix we show
that in the flower-pot model no bound states exist. Note that for an
interior Minkowskian observer the radius of the core is $\sigma a$. The
normalization coefficient $C_{l}$ is found from the condition (\ref%
{normcoefRl}):%
\begin{equation}
C_{l}^{2}=\frac{2\lambda J_{l+n/2}^{-2}(\lambda \sigma a)}{\pi ^{2}\sigma
\omega N(m_{k})\left[ \bar{J}_{\nu _{l}}^{2}(\lambda a)+\bar{Y}_{\nu
_{l}}^{2}(\lambda a)\right] },  \label{Clnorm}
\end{equation}%
with the barred notation for the cylindrical functions%
\begin{equation}
\bar{F}(z)\equiv zF^{\prime }(z)-\left[ \alpha _{\sigma }+z\frac{%
J_{l+n/2}^{\prime }(z\sigma )}{J_{l+n/2}(z\sigma )}\right] F(z)
\label{barredflow}
\end{equation}%
and%
\begin{equation}
\alpha _{\sigma }=\frac{1}{2}\left( 1-\frac{1}{\sigma }\right)
\left[ n-4\xi (n+1)\right] . \label{alfsig}
\end{equation}%
Note that $\bar{J}_{\nu _{l}}(\lambda a)=0$ for $\sigma =1$. Hence, the
parts in the Wightman function, in the VEVs of the field square and the
energy-momentum tensor due the non-trivial structure of the core in the
flower-pot model, are given by formulae (\ref{regWightout}), (\ref%
{regsquareout}) and (\ref{q1in}) respectively, where the tilted notations
for the modified Bessel functions are defined by (\ref{Barrednotmod}) with
the coefficient%
\begin{equation}
\mathcal{R}_{l}(a,z)=\alpha _{\sigma }+z\frac{I_{l+n/2}^{\prime }(z\sigma )}{%
I_{l+n/2}(z\sigma )}.  \label{Rcalflow}
\end{equation}%
For $\sigma =1$ one has $\tilde{I}_{\nu _{l}}(z)=0$ and as we could expect
the VEVs vanish. Using the value for the standard integral involving the
product of the functions $K_{\nu }$ given in Ref. \cite{Prud86}, in the case
of a massless scalar field the leading term for the asymptotic expansion
over $a/r$ can be presented in the form
\begin{equation}
\langle \varphi ^{2}\rangle _{c}\approx -\frac{\nu _{0}\Gamma (2\nu
_{0}+1/2)\Gamma (\nu _{0}+1/2)\mathcal{A}_{n}}{2^{2\nu _{0}+1}(a\sigma
)^{D-1}S_{D}\Gamma ^{3}(\nu _{0}+1)}\left( \frac{a}{r}\right) ^{2\nu
_{0}+D-1},  \label{phi2rgga}
\end{equation}%
where
\begin{equation}
\mathcal{A}_{n}=\frac{n\sigma -4\xi (n+1)(\sigma -1)-2\sigma \nu _{0}}{%
n\sigma -4\xi (n+1)(\sigma -1)+2\sigma \nu _{0}}.  \label{Acaln}
\end{equation}%
Note that for a minimally coupled scalar $\mathcal{A}_{n}=0$ and the
presented leading term vanishes. In figure \ref{fig1} we have plotted the
dependence of the part in the VEV of the field square induced by the core as
a function on the rescaled radial coordinate for minimally and conformally
coupled $D=3$ massless scalar fields in the flower-pot model with $\sigma
=0.5$

Now let us analyze the VEV of the energy-momentum tensor given by Eq. (\ref%
{q1in}) with the tilted notation given by (\ref{Barrednotmod}), (\ref%
{Rcalflow}). For large distances from the core, $r\gg a$, the main
contribution into the VEV of the energy-momentum tensor for a massless
scalar field comes from the $l=0$ summand. Under the assumption $\nu
_{0}\neq 0$ the leading terms of the asymptotic expansions have the form (no
summation over $i$)
\begin{equation}
\langle T_{i}^{k}\rangle _{c}\approx -\frac{2^{-2\nu _{0}}\sigma ^{1-D}%
\mathcal{A}_{n}\delta _{i}^{k}}{\pi \nu _{0}S_{D}\Gamma ^{2}(\nu _{0})a^{D+1}%
}\left( \frac{a}{r}\right) ^{2\nu _{0}+D+1}\int_{0}^{\infty }dzz^{2\nu
_{0}+2}F_{\nu _{0}}^{(i)}[K_{\nu _{0}}(z)].  \label{relfarDq}
\end{equation}%
The integrals in this formula can be evaluated using the value for the
integrals involving the product of the functions $K_{\nu }$ given in Ref.
\cite{Prud86}. As we see, for $\nu _{0}>0$ and for large distances from the
sphere the vacuum energy-momentum tensor is dominated by the part
corresponding to the point-like monopole.

\begin{figure}[tbph]
\begin{center}
\epsfig{figure=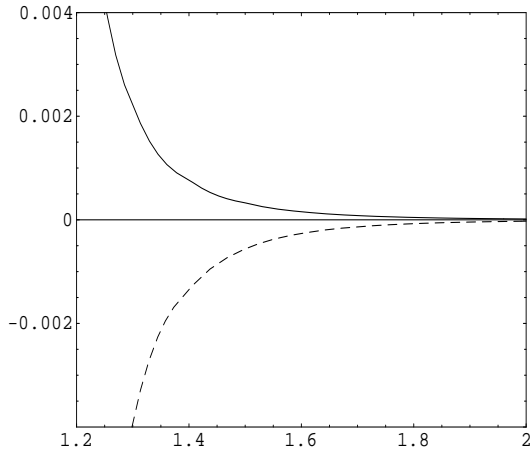,width=7.cm,height=6.cm}
\end{center}
\caption{The expectation value of the energy density, $a^{D+1}\langle
T_{0}^{0}\rangle _{c}$ induced in the region outside the core for $D=3$
massless scalar field as a function of $r/a$ in flower-pot model with $%
\protect\sigma =0.5$. The full/dashed curves correspond to
minimally/conformally coupled scalars.}
\label{fig2}
\end{figure}

As it has been mentioned above on the core surface the VEVs diverge. For the
region near the core the main contribution comes from large values of $l$.
By using the uniform asymptotic expansions for the modified Bessel functions
it can be seen that to the leading order $\langle \varphi ^{2}\rangle
_{c}\sim (r-a)^{2-D}$ and the components of the vacuum energy-momentum
tensor behave as $(r-a)^{-D}$ \ for the energy density and the azimuthal
stress and as $(r-a)^{1-D}$ for the radial stress. Due to surface
divergencies near the surface the total vacuum energy-momentum tensor is
dominated by the parts induced by the finite thickness of the core. As an
illustration, in figure \ref{fig2} we have presented the dependence of the
core-induced vacuum energy density as a function on the radial coordinate
for $D=3$ minimally and conformally coupled massless scalar fields in the
flower-pot model with $\sigma =0.5$.

\subsection{Interior region}

\label{subsec:flpotinter}

Now let us consider the vacuum polarization effects inside the core for the
flower-pot model. The corresponding eigenfunctions have the form given by
Eq. (\ref{phialf1}) with $f_{l}(r)=R_{l}(r,\lambda )$ and the function $%
R_{l}(r,\lambda )$ is defined by formula (\ref{Rlflow}). Substituting the
eigenfunctions into the mode sum formula for the corresponding Wightman
function one finds%
\begin{eqnarray}
\langle 0|\varphi (x)\varphi (x^{\prime })|0\rangle &=&\frac{2}{\pi
^{2}n\sigma S_{D}}\sum_{l=0}^{\infty }\frac{2l+n}{(\tilde{r}\tilde{r}%
^{\prime })^{n/2}}C_{l}^{n/2}(\cos \theta )  \notag \\
&&\times \int_{0}^{\infty }d\lambda \,\frac{\lambda J_{l+n/2}^{-2}(\lambda
\sigma a)}{\sqrt{\lambda ^{2}+m^{2}}}\frac{J_{l+n/2}(\lambda \tilde{r}%
)J_{l+n/2}(\lambda \tilde{r}^{\prime })}{\bar{J}_{\nu _{l}}^{2}(\lambda a)+%
\bar{Y}_{\nu _{l}}^{2}(\lambda a)}e^{i\sqrt{\lambda ^{2}+m^{2}}(t^{\prime
}-t)}.  \label{WFflowin}
\end{eqnarray}%
To find the renormalized VEVs of the field square and the energy-momentum
tensor we need to evaluate the difference between this function and the
corresponding function for the Minkowski bulk:%
\begin{equation}
\langle \varphi (x)\varphi (x^{\prime })\rangle _{\mathrm{sub}}=\langle
0|\varphi (x)\varphi (x^{\prime })|0\rangle -\langle 0_{M}|\varphi
(x)\varphi (x^{\prime })|0_{M}\rangle .  \label{WFsubinterior}
\end{equation}%
The appropriate form for the Minkowskian part is obtained from Eq. (\ref%
{Mink}) taking $\sigma =1$ and replacing $r\rightarrow \tilde{r}$. By using
the corresponding formula, for the subtracted Wightman function one finds%
\begin{eqnarray}
\langle \varphi (x)\varphi (x^{\prime })\rangle _{\mathrm{sub}} &=&\frac{2}{%
\pi ^{2}nS_{D}}\sum_{l=0}^{\infty }\frac{2l+n}{(\tilde{r}\tilde{r}^{\prime
})^{n/2}}C_{l}^{n/2}(\cos \theta )\int_{0}^{\infty }d\lambda \,\frac{\lambda
e^{i\sqrt{\lambda ^{2}+m^{2}}(t^{\prime }-t)}}{\sqrt{\lambda ^{2}+m^{2}}}
\notag \\
&&\times J_{l+n/2}(\lambda \tilde{r})J_{l+n/2}(\lambda \tilde{r}^{\prime })%
\left[ \frac{J_{l+n/2}^{-2}(\lambda \sigma a)/\sigma }{\bar{J}_{\nu
_{l}}^{2}(\lambda a)+\bar{Y}_{\nu _{l}}^{2}(\lambda a)}-\frac{\pi ^{2}}{4}%
\right] ,  \label{WFflowinsub}
\end{eqnarray}%
where the barred notation is defined by Eq. (\ref{barredflow}). The integral
in this formula is slowly convergent and the integrand is highly
oscillatory. In order to transform the expression for the subtracted
Wightman function into more convenient form, we note that the following
identity takes place%
\begin{equation}
\frac{1}{2}\sum_{s=1,2}\frac{C\{H_{l+n/2}^{(s)}(\sigma z),J_{\nu _{l}}(z)\}}{%
J_{l+n/2}(\sigma z)\bar{J}_{\nu _{l}}(z)}=1,  \label{identflow2new}
\end{equation}%
where we have introduced the notation%
\begin{equation}
C\left\{ f(\sigma z),g(z)\right\} =zf(\sigma z)g^{\prime }(z)-\left[ \alpha
_{\sigma }f(\sigma z)+zf^{\prime }(\sigma z)\right] g(z).  \label{Cfg}
\end{equation}%
Note that in terms of this notation one has
\begin{equation}
J_{l+n/2}(\sigma z)\bar{F}(z)=C\{J_{l+n/2}(\sigma z),F(z)\}.  \label{bartoC}
\end{equation}%
We add the left-hand side of Eq. (\ref{identflow2new}) with $z=\lambda a$ as
a coefficient to the term $\pi ^{2}/4$ in the square brackets of Eq. (\ref%
{WFflowinsub}). After this replacement the term in the square brackets is
written in the form%
\begin{eqnarray}
\frac{J_{l+n/2}^{-2}(\sigma z)/\sigma }{\bar{J}_{\nu _{l}}^{2}(z)+\bar{Y}%
_{\nu _{l}}^{2}(z)}-\frac{\pi ^{2}}{4} &=&\sum_{s=1,2}\frac{1}{%
2C\{J_{l+n/2}(\sigma z),J_{\nu _{l}}(z)\}}  \notag \\
&\times &\left[ \frac{1/\sigma }{C\{J_{l+n/2}(\sigma z),H_{\nu
_{l}}^{(s)}(z)\}}-\frac{\pi ^{2}}{4}C\{H_{l+n/2}^{(s)}(\sigma z),J_{\nu
_{l}}(z)\}\right] .  \label{identflow2}
\end{eqnarray}%
Note that both terms in the sum over $s$ on the right of this relation are
separately regular at the zeros of the function $C\{J_{l+n/2}(\sigma \lambda
a),J_{\nu _{l}}(\lambda a)\}$. Substituting (\ref{identflow2}) into formula (%
\ref{WFflowinsub}) we rotate the integration contour in the complex plane $%
\lambda $ by the angle $\pi /2$ for $s=1$ and by the angle $-\pi /2$ for $%
s=2 $. Under the condition $\tilde{r}+\tilde{r}^{\prime }+|t-t^{\prime
}|<2\sigma a$ the contribution from the semicircle with the radius tending
to infinity vanishes. Note that as we consider the points inside the core
this condition is satisfied in the coincidence limit. The integrals over the
segments $(0,im)$ and $(0,-im)$ of the imaginary axis cancel out and after
introducing the modified Bessel functions the subtracted Wightman function
can be presented in the form%
\begin{eqnarray}
\langle \varphi (x)\varphi (x^{\prime })\rangle _{\mathrm{sub}} &=&-\frac{1}{%
\pi nS_{D}}\sum_{l=0}^{\infty }\frac{2l+n}{(\tilde{r}\tilde{r}^{\prime
})^{n/2}}C_{l}^{n/2}(\cos \theta )\int_{m}^{\infty }dz\,\frac{zU_{l}(\sigma
,za)}{\sqrt{z^{2}-m^{2}}}  \notag \\
&&\times I_{l+n/2}(z\tilde{r})I_{l+n/2}(z\tilde{r}^{\prime })\cosh \!\left[
\sqrt{z^{2}-m^{2}}(t^{\prime }-t)\right] ,  \label{WFflowinsub1}
\end{eqnarray}%
with the notation%
\begin{equation}
U_{l}(\sigma ,z)=\frac{1/\sigma +C\{I_{l+n/2}(\sigma z),K_{\nu
_{l}}(z)\}C\{K_{l+n/2}(\sigma z),I_{\nu _{l}}(z)\}}{C\{I_{l+n/2}(\sigma
z),I_{\nu _{l}}(z)\}C\{I_{l+n/2}(\sigma z),K_{\nu _{l}}(z)\}}.
\label{Ulsigzet}
\end{equation}%
For points away from the core boundary the integral is exponentially
convergent in the coincidence limit and this formula is convenient for the
calculation of the VEVs of the field square and the energy-momentum tensor.

\begin{figure}[tbph]
\begin{center}
\epsfig{figure=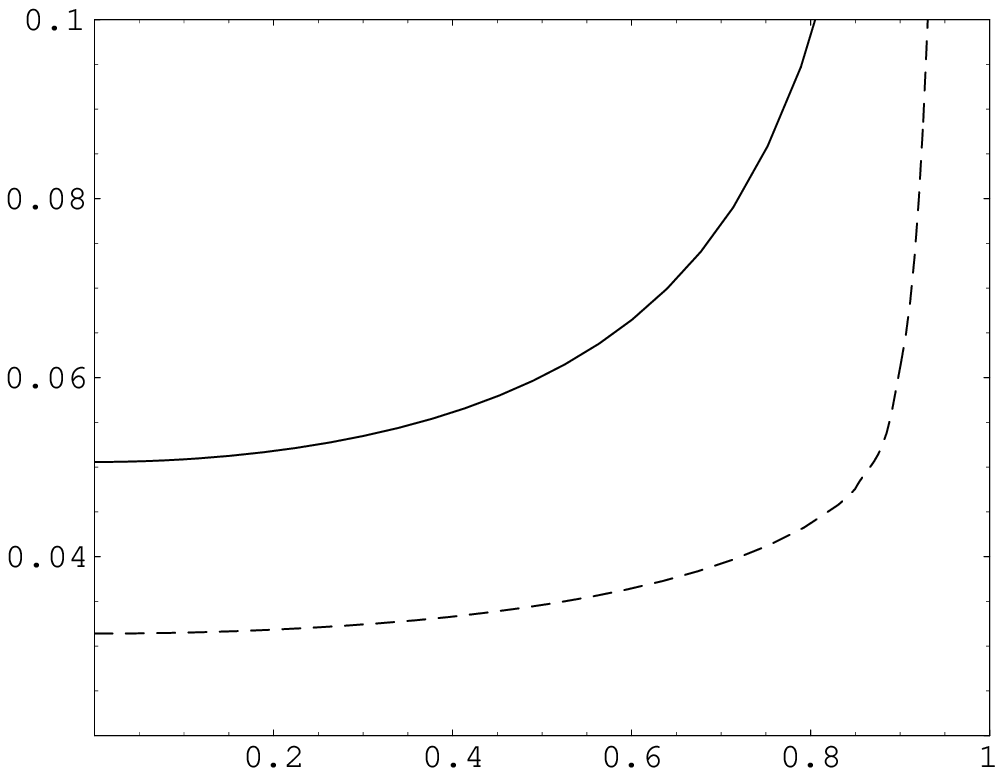,width=7.cm,height=6.cm}
\end{center}
\caption{The expectation value $a^{D-1}\langle \protect\varphi ^{2}\rangle _{%
{\mathrm{ren}}}$ inside the core for $D=3$ massless scalar field as a
function of $\tilde{r}/\protect\sigma a$ in the flower-pot model with $%
\protect\sigma =0.5$. The full/dashed curves correspond to
minimally/conformally coupled scalars.}
\label{fig3}
\end{figure}

Having the subtracted Wightman function we can evaluate the renormalized VEV
of the field square by taking the coincidence limit of the arguments in Eq. (%
\ref{WFflowinsub1}):%
\begin{equation}
\langle \varphi ^{2}\rangle _{\mathrm{ren}}=-\frac{1}{\pi S_{D}\tilde{r}^{n}}%
\sum_{l=0}^{\infty }D_{l}\int_{m}^{\infty }dz\frac{zU_{l}(\sigma ,za)}{\sqrt{%
z^{2}-m^{2}}}\,I_{l+n/2}^{2}(z\tilde{r}).  \label{phi2flowin}
\end{equation}%
So, by this result we can see that although the spacetime inside the core is
Minkowski one, there exists a vacuum polarization induced by the non-trivial
topology of the spacetime in the exterior region. In the limit $r\rightarrow
0$ the main contribution into Eq. (\ref{phi2flowin}) comes from the $l=0$
summand with the leading term
\begin{equation}
\langle \varphi ^{2}\rangle _{\mathrm{ren}}\approx -\frac{2^{1-D}}{\pi
^{D/2+1}\Gamma (D/2)}\int_{m}^{\infty }dz\,z^{D-1}\frac{U_{0}(\sigma ,za)}{%
\sqrt{z^{2}-m^{2}}},  \label{limitrto0}
\end{equation}%
and at the monopole center the renormalized VEV of the field square acquires
a non-vanishing regular contribution. For the points on the core surface the
VEV given by Eq. (\ref{phi2flowin}) diverges like $1/(a-r)^{D-2}$. Now let
us consider the limiting case $\sigma \ll 1$ under the fixed value $\sigma a$
which is the core radius for an internal Minkowskian observer. Introducing
in Eq. (\ref{phi2flowin}) a new integration variable $y=\sigma az$ and by
making use of the uniform asymptotic expansions for the modified Bessel
functions with the index $\nu _{l}$, we can see that to the leading order%
\begin{equation}
\langle \varphi ^{2}\rangle _{\mathrm{ren}}\approx -\frac{1}{\pi S_{D}\tilde{%
r}^{n}}\sum_{l=0}^{\infty }D_{l}\int_{m}^{\infty }dz\frac{zU_{l+n/2}(z\sigma
a)}{\sqrt{z^{2}-m^{2}}}\,\frac{I_{l+n/2}^{2}(z\tilde{r})}{%
I_{l+n/2}^{2}(z\sigma a)},  \label{phi2instro}
\end{equation}%
where we have introduced the notation%
\begin{equation}
U_{\nu }(y)=\frac{2\sqrt{\eta _{1}+y^{2}}}{\eta _{1}+y^{2}-\left[ \eta
_{2}+yI_{\nu }^{\prime }(y)/I_{\nu }(y)\right] ^{2}}-I_{\nu }(y)K_{\nu }(y),
\label{Unu}
\end{equation}%
with%
\begin{equation}
\eta _{1}=l\left( l+n\right) +n(n+1)\xi ,\;\eta _{2}=2\xi (n+1)-n/2.
\label{eta12}
\end{equation}%
Hence, in the limit $\sigma \rightarrow 0$ for a fixed radius of the core, $%
\sigma a$, the part in the renormalized VEV of the field square inside the
core tends to the finite limiting value. For large values of the mass,
assuming that $m(\sigma a-\tilde{r})\gg 1$, it can be seen that $\langle
\varphi ^{2}\rangle _{\mathrm{ren}}$ is suppressed by the factor $%
e^{-2m(\sigma a-\tilde{r})}$. In figure \ref{fig3} we have plotted the
renormalized VEV $\langle \varphi ^{2}\rangle _{\mathrm{ren}}$ inside the
core of the flower-pot model with $\sigma =0.5$ as a function of $\tilde{r}%
/\sigma a$ for minimally and conformally coupled massless scalars. Again we
can observe that there exists a strong dependence of this quantity on the
curvature coupling parameter.

\begin{figure}[tbph]
\begin{center}
\epsfig{figure=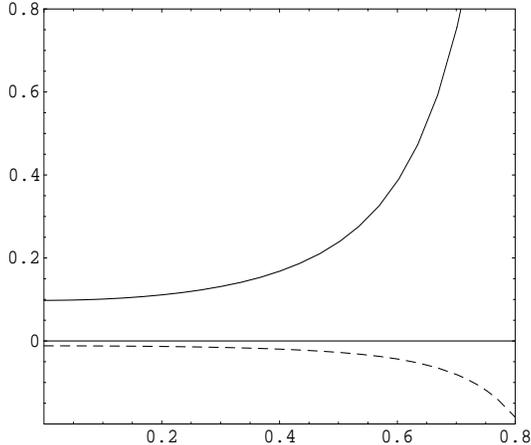,width=7.cm,height=6.cm}
\end{center}
\caption{The renormalized energy density, $a^{D+1}\langle T_{0}^{0}\rangle _{%
\text{ren}}$ inside the core for $D=3$ massless scalar field as a function
of $\tilde{r}/\protect\sigma a$ in the flower-pot model with $\protect\sigma %
=0.5$. The full/dashed curves correspond to minimally/conformally coupled
scalars.}
\label{fig4}
\end{figure}

The renormalized VEV of the energy-momentum tensor is found by using the
formula (\ref{mvevEMT}) with the subtracted Wightman functions. This leads
to the following formula (no summation over $i$)%
\begin{equation}
\langle T_{i}^{k}\rangle _{\mathrm{ren}}=-\frac{\delta _{i}^{k}}{2\pi S_{D}%
\tilde{r}^{n}}\sum_{l=0}^{\infty }D_{l}\int_{m}^{\infty }dz\frac{%
z^{3}U_{l}(\sigma ,za)}{\sqrt{z^{2}-m^{2}}}F_{l+n/2}^{(i)}[I_{l+n/2}(z\tilde{%
r})],  \label{emtflowin}
\end{equation}%
where the functions $F_{l+n/2}^{(i)}[f(y)]$ are defined by relations (\ref%
{Fineps})-(\ref{Finpeaz}) with the replacement $\nu _{l}\rightarrow l+n/2$.
At the core center the nonzero contribution to VEV (\ref{emtflowin}) comes
from the summands with $l=0$ and $l=1$ and one has
\begin{eqnarray}
\langle T_{0}^{0}\rangle _{\mathrm{ren}} &=&\frac{1}{2^{D}\pi ^{D/2+1}\Gamma
(D/2)}\int_{m}^{\infty }\frac{z^{D+1}dz}{\sqrt{z^{2}-m^{2}}}  \notag \\
&&\times \left[ \left( 4\xi +1-2\frac{m^{2}}{z^{2}}\right) U_{0}(\sigma
,za)+(4\xi -1)U_{1}(\sigma ,za)\right] ,  \label{encentre0} \\
\langle T_{1}^{1}\rangle _{\mathrm{ren}} &=&\langle T_{2}^{2}\rangle _{%
\mathrm{ren}}=\frac{1}{2^{D}\pi ^{D/2+1}D\Gamma (D/2)}\int_{m}^{\infty }%
\frac{z^{D+1}dz}{\sqrt{z^{2}-m^{2}}}  \notag \\
&&\times \left[ \left( \tilde{\xi}-2\right) U_{0}(\sigma ,za)+\tilde{\xi}%
U_{1}(\sigma ,za)\right] ,  \label{encentre}
\end{eqnarray}%
Note that for the conformally coupled massless scalar at the center one has $%
\langle T_{0}^{0}\rangle _{\mathrm{ren}}=-D\langle T_{1}^{1}\rangle _{%
\mathrm{ren}}$. This can also be obtained directly from the zero trace
condition. Near the core surface the components of the vacuum
energy-momentum tensor behave as $(a-r)^{-D}$ \ for the energy density and
the azimuthal stress and as $(a-r)^{1-D}$ for the radial stress. As in the
case of the field square, in the limit $\sigma \rightarrow 0$ for a fixed
radius of the core radius $\sigma a$, the part in the vacuum energy-momentum
tensor induced by the non-trivial core tends to the finite limiting value.
This limiting value is obtained from formula (\ref{emtflowin}) by making the
replacement $U_{l}(\sigma ,za)\rightarrow U_{l+n/2}(z\sigma
a)/I_{l+n/2}^{2}(z\sigma a)$. As in the case of the field square, for large
values of the mass for the field quanta the VEV $\langle T_{i}^{k}\rangle _{%
\mathrm{ren}}$ is exponentially suppressed by the factor $e^{-2m(\sigma a-%
\tilde{r})}$. The dependence of the renormalized interior vacuum energy
density on the radial coordinate is presented in figure \ref{fig4} for
minimally and conformally coupled massless scalar field in $D=3$ for the
geometry of a global monopole with $\sigma =0.5$.

\section{Conclusion}

\label{msec3:conc}

In the present paper we have considered the one-loop vacuum
effects for a massive scalar field with general curvature coupling
parameter on background of the $(D+1)$-dimensional global monopole
with non-trivial core structure. The previous papers on the
investigation of the vacuum polarization by the gravitational
field of the global monopole are concerned with the idealized
point-like model, where the curvature has singularity at the
origin. The exception is the Ref. \cite{Spin05}, where the vacuum
densities for a massless scalar field are studied outside the
monopole core with the interior de Sitter geometry. Here we
consider the general spherically symmetric static model of the
core with finite thickness, described by the line element
(\ref{metricinside}), and investigate the vacuum properties in
both exterior and interior regions. Among the most important
characteristics of these properties, which carry an information
about the core structure, are the VEVs for the field square and
the energy-momentum tensor.  In order to obtain these expectation
values we first construct the positive frequency Wightman
function. In the region outside the core this function is
presented as a sum of two distinct contributions. The first one
corresponds to the Wightman function for the geometry of a
point-like global monopole and the second one is induced by the
non-trivial structure of the monopole's core. The latter is given
by formula (\ref{regWightout}), where the tilted notation is
defined by formula (\ref{Barrednotmod}) with the coefficient from
(\ref{Rlcal}) for the model without an infinitely thin spherical
shell on the boundary of the core. This coefficient is determined
by the radial part of the interior eigenfunctions and describes
the influence of the core properties on the vacuum characteristics
in the exterior region. In the case of the core model with a thin
shell on the boundary the derivatives of the metric tensor
components are discontinuous on the core surface. This leads to
the delta function type contribution to the Ricci scalar and,
hence to the equation for the radial eigenfunctions in the case of
the non-minimally coupled scalar field. As a result, the radial
eigenfunctions have a discontinuity in their slope at the core
boundary. This leads to an additional term in the coefficient of
the tilted notation which is
proportional to the trace of the surface energy-momentum tensor (see Eq. (%
\ref{newRlcal})).

By using the formula for the Wightman function, in section %
\ref{sec:phi2emt} we have investigated the influence of the non-trivial core
structure on the VEVs of the field square and the energy-momentum tensor. As
in the exterior region the local geometry is the same as that in the
point-like global monopole model, the presence of the core does not lead to
additional divergences for the points outside the core. As a result, the
parts in these VEVs induced by the core are directly obtained from the
corresponding part of the Wightman function for the case of the field square
and by applying on this function a certain second-order differential
operator and taking the coincidence limit for the energy-momentum tensor.
These parts are given by formulae (\ref{regsquareout}) and (\ref{q1in}) for
the field square and the energy-momentum tensor respectively. They diverge
as the boundary of the core is approached. The surface divergences in the
VEVs of the local observables are well--known in quantum field theory with
boundaries and are investigated for various boundary geometries. We have
investigated the asymptotic behavior of the core induced VEVs near the core
boundary and at large distances from the core. In particular, at large
distances and for a massless scalar field with $\nu _{0}>0$, the ratio of
the core induced and the point-like monopole parts decay as $(r/a)^{2\nu
_{0}}$ for the both field square and the energy-momentum tensor. For the
special case with $\nu _{0}=0$ this ratio decays logarithmically and
long-range effects of the monopole's core appear. In the limit of strong
gravitational fields corresponding to small values of the parameter $\sigma $%
, the behavior of the core induced parts is completely different for
minimally and non-minimally coupled fields. The corresponding VEVs are
suppressed by the factor $\exp [-(2/\sigma )\sqrt{n(n+1)\xi }\ln (a/r)]$ for
the non-minimally coupled scalar and behave like $\sigma ^{1-D}$ for the
minimally coupled field.

As an example of the application of the general results, in section \ref%
{sec:flowerpot} we have considered a simple core model with a flat spacetime
inside the core, so called flower-pot model. The corresponding surface
energy-momentum tensor on the boundary of the core is obtained from the
matching conditions and has the form given by Eq. (\ref{surfemtflow}). The
core induced parts of the exterior VEVs in this model are obtained from the
general results by taking the function in the coefficient of the tilted
notation form Eq. (\ref{Rcalflow}). For the flower-pot model we have also
investigated the vacuum densities inside the core. Though the spacetime
geometry inside the core is Monkowskian, the non-trivial topology of the
exterior region induces vacuum polarization effects in this region as well.
In order to find the corresponding renormalized VEVs of the field square and
the energy-momentum tensor we have derived a closed formula, Eq. (\ref%
{WFflowinsub1}), for the difference of the interior Wightman function and
the Wightman function for the Minkowski spacetime. The subtracted function
is finite in the coincidence limit and can be directly used for the
evaluation of the VEVs of the field square and the energy-momentum tensor.
The latter quantities are given by formulae (\ref{phi2flowin}) and (\ref%
{emtflowin}). As in the case of the exterior region, we have considered
various limiting cases when the general formulae are simplified. In
particular, we have shown that in the limit $\sigma \ll 1$ under the fixed
value $\sigma a$, which is the core radius for an internal Minkowskian
observer, the renormalized VEVs of the field square and the energy-momentum
tensor tend to finite limiting values.

\section*{Acknowledgement}

AAS was supported by PVE/CAPES Program and in part by the Armenian Ministry
of Education and Science Grant No. 0124. ERBM thanks Conselho Nacional de
Desenvolvimento Cient\'\i fico e Tecnol\'ogico (CNPq) and FAPESQ-PB/CNPq
(PRONEX) for partial financial support.

\appendix

\section{Contribution of bound states}

\label{sec:app1}

In this appendix we consider the contribution of possible bound states into
the VEVs. For this states the quantity $\lambda $ is purely imaginary, $%
\lambda =i\eta $, and the corresponding radial eigenfunction in the region $%
r>a$ is the function $A_{bl}r^{-n/2}K_{\nu _{l}}(\eta r)$ with the
normalization coefficient $A_{bl}$. To have a stable ground state we will
assume that $\eta <m$. From the continuity of the eigenfunctions at $r=a$
one has%
\begin{equation}
R_{l}(a,i\eta )=A_{bl}a^{-n/2}K_{\nu _{l}}(\eta a),  \label{appcond1}
\end{equation}%
and from the continuity of the radial derivative we see that the possible
bound states are solutions of the equation%
\begin{equation}
\tilde{K}_{\nu _{l}}(\eta a)=0,  \label{appBSeq}
\end{equation}%
with the notation from (\ref{Barrednotmod}). The normalization condition for
the bound states is as follows:%
\begin{equation}
\int_{r_{0}}^{a}dr\,e^{-u+v+(D-1)w}R_{l}^{2}(r,i\eta )+A_{bl}\sigma
^{D-1}\int_{a}^{\infty }dr\,rK_{\nu _{l}}^{2}(\eta r)=\frac{1}{2\omega
N(m_{k})},  \label{appnormcond}
\end{equation}%
from which the normalization constant can be found. In order to evaluate the
integrals in this formula we note that for the solution $f_{\omega l}(r)$ to
radial equation (\ref{fleq}) the following formula takes place
\begin{equation}
\int dr\,e^{-u+v+(D-1)w}f_{\omega l}(r)f_{\omega _{1}l}(r)=\frac{%
e^{-u+v+(D-1)w}}{\omega _{1}^{2}-\omega ^{2}}\left[ f_{\omega l}^{\prime
}(r)f_{\omega _{1}l}(r)-f_{\omega l}(r)f_{\omega _{1}l}^{\prime }(r)\right] .
\label{appIntForm}
\end{equation}%
In particular, in the limit $\omega _{1}\rightarrow \omega $ one finds%
\begin{equation}
\int dr\,e^{-u+v+(D-1)w}f_{\omega l}^{2}(r)=\frac{e^{-u+v+(D-1)w}}{2\omega }%
\left[ f_{\omega l}^{\prime }(r)\frac{\partial }{\partial \omega }f_{\omega
l}(r)-f_{\omega l}(r)\frac{\partial }{\partial \omega }f_{\omega l}^{\prime
}(r)\right] .  \label{appintform1}
\end{equation}%
Applying to the integrals in Eq. (\ref{appnormcond}) this formula and using
the continuity of the radial eigenfunctions at $r=a$, for the normalization
coefficient one finds%
\begin{equation}
A_{bl}^{2}=-\frac{\eta \sigma ^{1-D}\tilde{I}_{\nu _{l}}(\eta a)}{\omega
N(m_{k})(\partial /\partial \eta )\tilde{K}_{\nu _{l}}(\eta a)}.
\label{appCl}
\end{equation}%
To obtain this formula we have used the relation%
\begin{equation}
K_{\nu _{l}}(\eta a)=1/\tilde{I}_{\nu _{l}}(\eta a),  \label{appKItild}
\end{equation}%
valid for the solutions of Eq. (\ref{appBSeq}). As a result, for the
contribution of the bound state with $\lambda =i\eta $ to the Wightman
function we have the formula%
\begin{eqnarray}
\langle \varphi (x)\varphi (x^{\prime })\rangle _{\mathrm{bs}} &=&-\frac{%
\sigma ^{1-D}}{nS_{D}}\sum_{l=0}^{\infty }\frac{2l+n}{(rr^{\prime })^{n/2}}%
C_{l}^{n/2}(\cos \theta )  \notag \\
&&\times \frac{\eta \tilde{I}_{\nu _{l}}(\eta a)}{(\partial /\partial \eta )%
\tilde{K}_{\nu _{l}}(\eta a)}\frac{e^{i(t^{\prime }-t)\sqrt{m^{2}-\eta ^{2}}}%
}{\sqrt{m^{2}-\eta ^{2}}}K_{\nu _{l}}(\eta r)K_{\nu _{l}}(\eta r^{\prime }).
\label{appWFbs}
\end{eqnarray}%
In the case when several bound states are present the sum of their separate
contributions should be taken. Now the Wightman function is the sum of the
part coming from the modes with real $\lambda $ given by Eq. (\ref{extdif})
and of the part coming from the bound states given by Eq. (\ref{appWFbs}).
In order to transform the first part we again rotate the integration contour
in Eq. (\ref{extdif}) by the angle $\pi /2$ for $s=1$ and by the angle $-\pi
/2$ for $s=2$. But now we should take into account that the integrand has
poles at $\lambda =\pm i\eta $ which are zeroes of the functions $\bar{H}%
_{\nu _{l}}^{(s)}(\lambda a)$ in accordance with Eq. (\ref{appBSeq}).
Rotating the integration contour we will assume that the pole $(-1)^{s}i\eta
$, $s=1,2$, on the imaginary axis is avoided by the semicircle $C_{\rho
}^{(s)}$ in the right half plane with small radius $\rho $ and with the
center at this pole. The integration over these semicircles will give an
additional contribution%
\begin{equation}
-\frac{\sigma ^{1-D}}{4nS_{D}}\sum_{l=0}^{\infty }\frac{2l+n}{(rr^{\prime
})^{n/2}}C_{l}^{n/2}(\cos \theta )\sum_{s=1}^{2}\int_{C_{\rho
}^{(s)}}d\lambda \,\lambda \frac{e^{i\sqrt{\lambda ^{2}+m^{2}}(t^{\prime
}-t)}}{\sqrt{\lambda ^{2}+m^{2}}}\frac{\bar{J}_{\nu _{l}}(\lambda a)}{\bar{H}%
_{\nu _{l}}^{(s)}(\lambda a)}H_{\nu _{l}}^{(s)}(\lambda r)H_{\nu
_{l}}^{(s)}(\lambda r^{\prime }).  \label{appbscont}
\end{equation}%
By evaluating the integrals in this formula it can be seen that this term
exactly cancels the contribution (\ref{appWFbs}) coming from the
corresponding bound state (for the similar cancellation in the Casimir
effect with Robin boundary condition see Ref. \cite{Rome02}). Hence, we
conclude that the formulae given above for the core induced parts in the
VEVs are valid in the case of the presence of bound states as well.

In order to see the possibility for the appearance of bound states in the
flower-pot model, we note that introducing a function $F_{l}(r)=r_{\ast
}^{(D-1)/2}f_{l}(r)$ with $r_{\ast }=r$ outside the core and $r_{\ast }=%
\tilde{r}$ inside the core, equation (\ref{fleq}) for the radial part of the
eigenfunctions is written in the form of the Schrodinger equation. The
corresponding effective potential is equal $[\nu _{l}^{2}+(n-1)/4]/\sigma
^{2}r^{2}$ in the exterior region and $[l(l+n)+(n^{2}-1)/4]/\tilde{r}^{2}$
in the interior region. Under the conditions $\nu _{l}^{2}\geqslant 0$ and $%
n>0$ assumed earlier, the potential is non-negative and, hence, in the
flower-pot model no bound states exist.

\end{document}